\documentclass[prb,aps,twocolumn,showpacs,preprintnumbers,boldmath,amsmath,amssymb,floatfix]{revtex4-1}
\usepackage{times}
\usepackage{graphicx}
\usepackage{amssymb}
\usepackage{amsmath}
\usepackage{amsfonts}
\usepackage{dcolumn}
\usepackage{braket}
\usepackage{bm}
\usepackage{multirow}
\usepackage[colorlinks=true,
            linkcolor=red,
            urlcolor=blue,
            citecolor=blue]{hyperref}
\usepackage{amstext}
\usepackage{framed}
\DeclareGraphicsExtensions{.pdf,.eps,.jpg}

\begin{document}
\title{Site-substitution in GdMnO$_3$ : effects on structural, electronic and magnetic properties}

\author{Sudipta Mahana$^1$$^,$$^2$$^,$$^\dagger$, Bipul Rakshit$^3$, Pronoy Nandi$^1$$^,$$^2$,Raktima Basu$^4$, Sandip Dhara$^4$, U. Manju$^5$ , Subhendra D. Mahanti$^6$ and D. Topwal$^1$$^,$$^2$$^,$}

\email {dinesh.topwal@iopb.res.in, dinesh.topwal@gmail.com}

\affiliation{$^1$Institute of physics, Sachivalaya Marg, Bhubaneswar - 751005, India\\
$^2$Homi Bhabha National Institute, Training School Complex, Anushakti Nagar, Mumbai - 400085, India\\
$^3$Center for Superfunctional Materials, Ulsan National Institute of Science and Technology, Ulsan - 44919, South Korea\\
$^4$Surface and Nanoscience Division, Indira Gandhi Centre for Atomic Research, HBNI, Kalpakkam - 603102, India\\
$^5$ CSIR -Institute of Minerals and Materials Technology, Bhubaneswar - 751013, India\\
$^6$ Department of Physics and Astronomy, Michigan State University, East Lansing, Michigan 48824, USA}

\begin{abstract}
We report on detailed structural, electronic and magnetic studies of GdMn$_{1-x}$Cr$_x$O$_3$ for Cr doping levels 0 $\le$ $x$ $\le$ 1. In the solid solutions, the Jahn-Teller distortion associated with Mn$^{3+}$ ions gives rise to major changes in the ${bc}$-plane sub-lattice and also the effective orbital ordering in the ${ab}$-plane, which persist up to the compositions $x$ $\sim$ 0.35. These distinct features in the lattice and orbital degrees of freedom are also correlated with $bc$-plane anisotropy of the local Gd environment. A gradual evolution of electronic states with doping is also clearly seen in O $K$-edge x-ray absorption spectra. Evidence of magnetization reversal in field-cooled-cooling mode for $x$ $\ge$ 0.35 coinciding the Jahn-Teller crossover, suggests a close correlation between magnetic interaction and structural distortion. These observations indicate a strong entanglement between lattice, spin, electronic and orbital degrees of freedom. The nonmonotonic variation of remnant magnetization can be explained by doping induced modification of  magnetic interactions. Density functional theory  calculations are consistent with a layer-by-layer type doping with ferromagnetic (antiferomagnetic) coupling between Mn (Cr) ions for intermediate compound ($x$ = 0.5), which is distinct from that observed for the end members GMnO$_3$ and GdCrO$_3$. 
\end{abstract}

\maketitle

\section{Introduction}
Functional oxides with perovskite structures (${AB}$O$_3$) are very active research area not only due to their potential technological applications but also for their fundamental importance in basic scientific research. An unusual aspect of perovskites is their ability to incorporate almost every element of the periodic table at the $A$ and $B$ sites due to their capacity to accommodate various structural distortions \cite{Mitchell}. External parameters like temperature, pressure and chemical compositions, can also drive such distortions, which leads to an extraordinary richness of physical properties within the family of perovskites. Structural distortions in perovskites are mainly associated with three main features with respect to their ideal cubic structure:\cite{Mitchell,glazer1972classification,glazer1975simple} (i) rotation (tilt) of $B$O$_6$ octahedra, (ii) polar cation displacements, which often lead to ferroelectricity, and (iii) distortions of the octahedra, such as the Jahn-Teller (JT) distortion. \\
The rare-earth manganites ($R$MnO$_3$) invoked great interest owing to the JT character of  Mn$^{3+}$ ions (${t_{2g}^3}$${e_g^1}$ ), exhibiting orbital ordering along with highly anisotropic Mn-O bond lengths \cite{zhou2006unusual}. A complex interplay among the spin, orbital and lattice degrees of freedom has led to a large number of intriguing physical properties in $R$MnO$_3$ such as colossal magnetoresistance \cite{rao1998colossal}, charge and orbital ordering \cite{mori1999microstructure,van1999charge,asaka2002charge}, metal-insulator transition \cite{kawano1997magnetic,fukumoto1999microscopic}, complex spin structures \cite{mochizuki2009microscopic},  multiferroic properties with significant magnetoelectric coupling \cite{cheong2007multiferroics}. In contrast to Mn$^{3+}$, Cr$^{3+}$ is JT inactive ion because of having completely empty $e_g$ orbitals and therefore the oxygen octahedra are more regular. However, most of the members of $R$CrO$_3$ have been reported to be multiferroic materials at considerable high temperature \cite{rajeswaran2012field,mahana2017local}. Additionally, $R$CrO$_3$ systems are of great interest as these exhibit complex magnetic properties such as spin-reorientation (SR), spin-flipping (SF) and temperature induced magnetization reversal (TMR) etc. \cite{mahana2017complex,cao2014magnetization,el2014local}\\ 

GdMnO$_3$ with the orthorhombic ${Pbnm}$ structure, exhibits incommensurate sinusoidal magnetic structure arising from competing nearest-neighbor ferromagnetic (NN-FM) and next-nearest-neighbor antiferromagnetic (NNN-AFM) interaction followed by a canted-$A$-type ordering in the Mn-sublattice \cite{mochizuki2009microscopic,mahana2017giant}. Additionally, a low temperature ferroelectric ordering is established, caused by Gd$^{3+}$-Mn$^{3+}$ spin interactions and/or lattice distortions associated with magnetic field-induced spin rearrangements \cite{kimura2005magnetoelectric,moreira2012magnetically}. GdCrO$_3$ is one of the $G$-type antiferromagnetic (AFM) $R$CrO$_3$ compounds, exhibiting extremely rich magnetic properties like TMR, SF, SR and others \cite{mahana2017complex}. It has non-centrosymmetric ${pna2_1}$ structure, associated with the ferroelectric transition concurrent to Cr magnetic ordering temperature with significant magnetoelectric coupling \cite{rajeswaran2012field,mahana2017local}. Although the parent compounds without doping are well investigated, the doped solid solution GdMn$_{1-x}$Cr$_x$O$_3$ is largely unexplored \cite{modi2015structural}. Various interesting properties have been reported in similar type of mixed cation compositions such as DyMn$_{1-x}$Fe$_x$O$_3$ \cite{chiang2011effect}, LaMn$_{1-x}$Fe$_x$O$_3$ \cite{long2013structural}, TbMn$_{1-x}$Fe$_x$O$_3$ \cite{fang2016observation}, YbMn$_{1-x}$Fe$_x$O$_3$ \cite{huang2007structural}, TbMn$_{1-x}$Cr$_x$O$_3$ \cite{staruch2014evidence} and others. This has motivated us to investigate the GdMn$_{1-x}$Cr$_x$O$_3$ series.\\

In this paper we present systematic structural, electronic and magnetic investigations of the solid solutions, GdMn$_{1-x}$Cr$_x$O$_3$ (0 $\le$ $x$ $\le$ 1). Doping Gd$M$O$_3$ ($M$ = Mn/Cr) gives rise to $M$-valence mixing and local static distortions around the doping ion, resulting in average lattice distortions in the compound. However, at considerable Cr-doping, the distortion is negligible resulting in a crossover from JT active region to JT inactive region.

 \section{Experimental and Theoretical Details}
Polycrystalline samples of Cr-doped gadolinium manganites, GdMn$_{1-x}$Cr$_x$O$_3$ (0 $\le$ $x$ $\le$ 1) were prepared by solid-state synthesis technique as reported elsewhere \cite{mahana2017giant}. The crystalline structure and phase purity of the solid solutions were confirmed by x-ray diffraction (XRD) measurements using Bruker D8 Advance X-ray diffractometer equipped with Cu $K_{\alpha}$ radiation. Rietveld refinements of the obtained powder XRD patterns were carried out using the FULLPROF program. Raman spectroscopy measurements were performed using a micro-
Raman spectrometer (inVia, Renishaw, United Kingdom) with 514.5 nm excitation of an Ar$^+$ laser. Spectra were collected in the backscattering configuration using a thermoelectrically cooled CCD camera as the detector and a long working distance 50$\times$ objective with a numerical aperture of 0.45 was used for
the acquisition. Magnetization  measurements were carried out using SQUID-VSM from Quantum Design US. The O $K$-edge x-ray absorption
spectra were recorded in total electron yield (TEY) mode at the CIRCULARPOLARIZATION beamline at the Elettra synchrotron radiation facility. \\

 Our theoretical calculations of the structural, electronic and magnetic properties were based on density functional theory, using generalized gradient approximation (GGA) with Perdew Burke Ernzerhof  for solids (PBEsol) \cite{perdew2008restoring} parameterization for the exchange correlation potential, the projector argumented wave (PAW) method \cite{kresse1999ultrasoft}, and a plane-wave basis set, as implemented in the Vienna ab-initio simulation package (VASP) \cite{kresse1996efficient}. The interaction between ions and electrons was approximated with PAW potentials, treating 3$p$, 3$d$ and 4$s$ for Cr/Mn and 2$s$ and 2$p$ for O as valence electrons. For Brillouin zone sampling, we chose 12{$\times$}12{$\times$}8 Monkhorst-Pack $k$-point mesh \cite{monkhorst1976special} and the wave-function was expanded in a basis set consisting of plane waves with kinetic energies less than or equal to 770 eV. Using these parameters, an energy convergence of less than 1 meV/formula unit (f.u.) was achieved. Structures were fully relaxed until residual Hellmann–Feynman (HF) forces were smaller than 0.001 eV/$\AA{}$ while maintaining the symmetry constraints of the given space group. Gd 4$f$ electrons were treated as valence electrons for parent compounds. We performed calculations using different Hubbard $U$ values up to 4 eV for Mn/Cr and 4 eV for Gd and results for $U$ =  3 eV for Mn/Cr and 4 eV for Gd explained satisfactorily the experimental results, which would be explained in later section. In GdMn$_{0.5}$Cr$_{0.5}$O$_3$, Gd 4$f$ electrons were assumed as core electrons to reduce the calculations time. Irrespective of this, the 4$f$ states lie deep in energy and they are almost completely localized so that they do not affect other valence states \cite{yamauchi2008magnetically}. To visualize the orbital ordering in GdMnO$_3$, in addition to the global $X$, $Y$, $Z$ orthorhombic frame a local frame specific to each Jahn-Teller-type distorted MnO$_6$ octahedron was defined choosing $x$, $y$, $z$ along the middle, short, and long Mn-O axes, respectively \cite{yamauchi2008magnetically}. 

\section{Results and discussion}
 Figure 1(a) depicts room temperature XRD patterns of the solid solutions GdMn$_{1-x}$Cr$_x$O$_3$ (0 $\le$ $x$ $\le$ 1) along with the corresponding Rietveld refined patterns considering orthorhombic ${Pbnm}$ space group except for GdCrO$_3$, which is fitted well using ${Pna2_1}$ space group \cite{mahana2017local}. The composition dependent evolution of the lattice parameters ($a$, $b$, and $c$/$\sqrt{2}$ ) and cell volume ($V$) in GdMn$_{1-x}$Cr$_x$O$_3$ are shown in Fig. 1(b). A remarkable decrease in the value of $b$-axis, accompanied with an increase of the $c$-axis value has been observed with increasing Cr-content, while the $a$-axis remains almost constant. This suggests that lattice degrees of freedom confined to the ${bc}$-plane are strongly affected by the substitution of Cr. Further, the decrease in the cell volume reveals that the $c$-axis elongation is dominated by the $b$-axis reduction. Such structural characteristics can not be explained by considering the ionic radii of Cr$^{3+}$ ion (0.615 \AA{}) in place of Mn$^{3+}$ ion (0.645 \AA{}) alone.  \\
\begin{figure}[!ht]
 \centering
 \includegraphics[height=12cm,width=9.5cm]{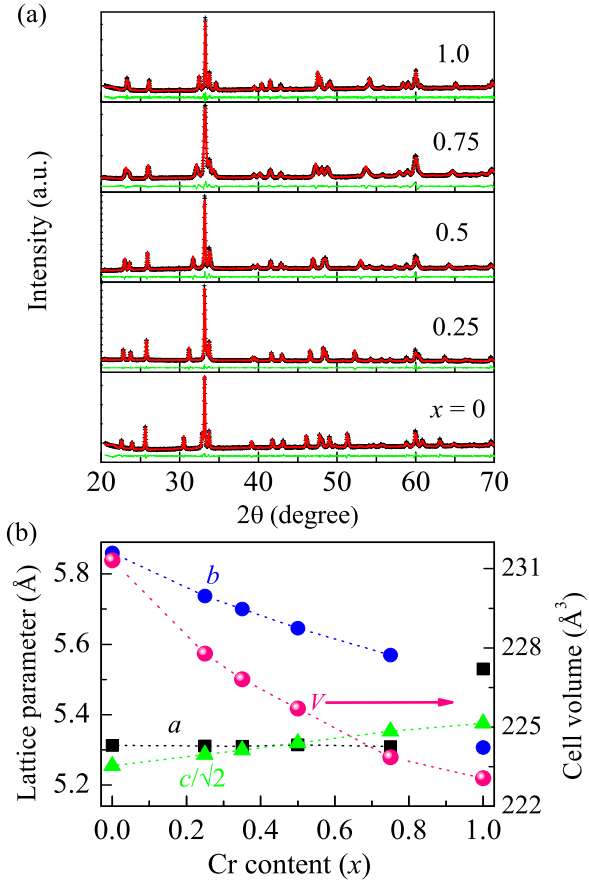}
\caption{(Color online) (a) The Rietveld-refinement plots of room temperature XRD patterns of GdMn$_{1-x}$Cr$_x$O$_3$ ($\emph{x}$ = 0, 0.25, 0.5, 0.75 and 1.0) in the space group of $Pbnm$. Experimental data is presented with symbol while the fitted curves from the Rietveld analysis are represented as red line. Difference spectra (difference between experimental data and fitting) is  plotted in blue line. (b) Evolution of
the cell parameters ($a$, $b$, and, c/$\sqrt2$ (left panel) and cell volume ($V$) (right panel) as a function of compositions ($x$). As GdCrO$_3$ has ${pna2_1}$ symmetry, lattice parameters $a$ and $b$ interchanges with respect to other compositions having ${Pbnm}$ symmetry. }
 \label{fig1}
 \end{figure}
Hence a detailed Reitveld refinement of the XRD patterns were carried out and the composition-dependent variations of three $M$-O bonds ($M$ = Mn/Cr) in the $M$O$_6$ octahedra, obtained from Reitveld refinements are shown in Fig. 2 (a), with $l_x$, $l_y$ and $l_z$ denoting bond lengths along the respective local axes discussed in experimental and theoretical details. The intrinsic octahedral distortion in the orthorhombic structure allows the short and long bonds to lie within the $ab$-plane and the intermediate bond length along the $c$-axis \cite{chiang2011effect,zhou2010intrinsic}. The large differences among the three $M$-O bond-lengths in GdMnO$_3$ are correlated with the cooperative JT distortion of Mn$^{3+}$ ion along with a contribution from intrinsic structural distortion. In contrast, GdCrO$_3$ exhibits a regular structure with similar bond lengths of $l_x$, $l_y$ and $l_z$, consistent with the quenched JT distortion for Cr$^{3+}$ ion. The local modes characterizing the JT distortion are defined as  in-(${ab}$) plane orthorhombic distortion, $Q_2$ [ = $l_y$-$l_x$] and out-of-plane tetragonal-like distortions, $Q_3$ [ = (2$l_z$-$l_x$-$l_y$)/$\sqrt{3}$] \cite{chiang2011effect,zhou2008intrinsic,tachibana2007jahn}, which are illustrated in Fig. 2 (b). The large positive value of $Q_2$ in GdMnO$_3$ is associated with the cooperative JT distortion, which is along the $b$-axis and ${Q_3}$ with negative sign indicates that an out-of-plane distortion along the $c$-axis is competing with the JT distortion \cite{chiang2011effect,zhou2008intrinsic}. This implies that the lattice deformation is primarily confined to the ${bc}$-plane sub-lattice. In addition, the larger magnitude of $Q_2$ over $Q_3$ also indicates that the increase of $c$-axis is largely overwhelmed by decrease of $b$-axis. Upon approaching towards GdCrO$_3$ the decrease of  both ${Q_2}$ and ${Q_3}$ indeed reveal gradual decrease of both JT and tetragonal distortions. An intriguing slope crossover in both ${Q_2}$ and ${Q_3}$ around $x$ $\sim$ 0.35 (guided by the dotted lines in ${Q_2}$) indicates the suppression of long range structural distortion associated with local JT distortion at the Mn sites. \\
\begin{figure*}[!ht]
 \centering
 \includegraphics[height=12cm,width=15cm]{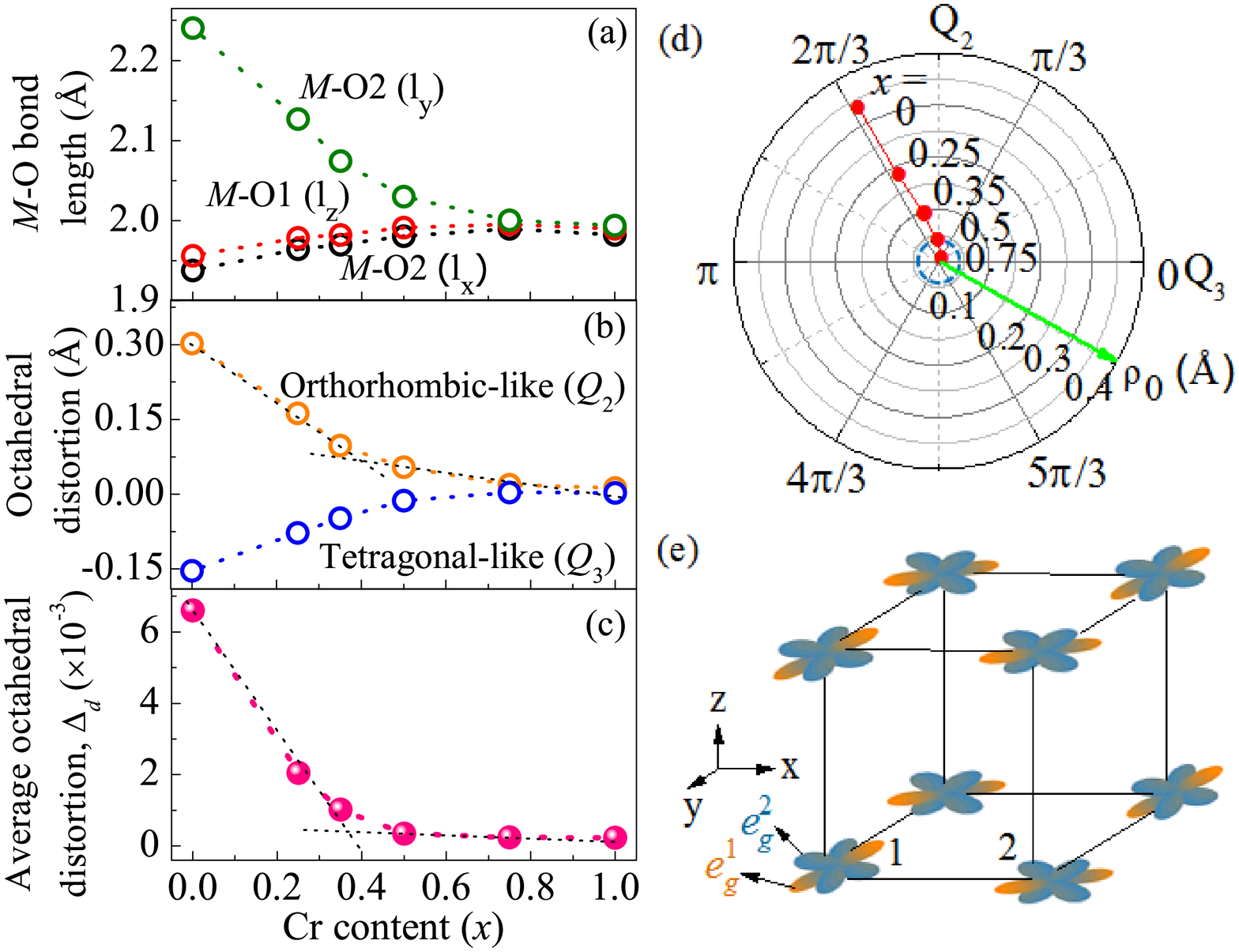}
\caption{(Color online) (a) The composition-dependent variations
of the $M$-O bonds in the $M$O$_6$ octahedra of GdMn$_{1-x}$Cr$_x$O$_3$, with the long, short $M$-O2 bonds and the middle $M$-O1 bond, respectively. O1 (O2) represents the apical (equatorial) oxygen along the $c$-axis. (b) The composition-dependent variations of in-($ab$)plane orthorhombic-like ($Q_2$) and out-of-plane tetragonal-like ($Q_3$) distortions . Dotted line guiding $Q_2$ point to the slope crossover  around $x$ $\sim$ 0.35. (c) Variation of average octahedral distortion evaluated by $\Delta_d$ . Dotted lines guiding $\Delta_d$ to the slope crossover as a result of effective supression of JT orbital ordering around $x$ $\sim$ 0.35. All the parameters are derived from the Rietveld refinements of the respective XRD patterns. (d) The polar plot of $\rho_0$ (= ${Q_2^2}$ + ${Q_3^2}$) and $\phi$ (= tan$^{-1}$($\frac{Q_3}{Q_2}$), which are used to describe the orbital mixing in GdMn$_{1-x}$Cr$_x$O$_3$. (e) Schematic diagram of $e_g$ orbitals of Mn$^{3+}$ due to the JT orbital ordering.}
 \label{fig2}
 \end{figure*}
Further examination of the average octahedral distortion $\Delta$$_d$  [= (1/6) $\sum$$_{n=1-6}$ [(${d_n}$-$\braket{d}$)/$\braket{d}$]$^2$, where ${d_n}$ ($\braket{d}$) is the individual (average) $M$-O bond length] (depicted in Fig. 2 (c)), shows shows a slope changeover around $x$ $\sim$ 0.35 (guided by the dotted lines), a characteristic of crossover from JT-active region to JT-inactive region.\\ 

JT-effect results in the lifting of degeneracy of e$_g$ orbitals of Mn$^{3+}$ ions and building up orbital ordering in the material. Thus, the JT distortion is conjugated with the $ab$-plane staggered orbital ordering. A polar plot of  magnitude of the octahedral-site distortion, $\rho_0$ (= ${Q_2^2}$ + ${Q_3^2}$) versus the angle $\phi$ (= tan$^{-1}$($\frac{Q_3}{Q_2}$) was mapped for the compositions as shown in Fig. 2 (d), where $\phi$ opens from the ${Q_2}$ axis in anticlockwise direction \cite{kanamori1960crystal}.  \\

The description of the $e_g$ orbital associated with the $M$ atom in an $M$O$_6$ octahedron can be made by the wave function $\psi$ with a linear combination of orbitals $\ket{x^2-y^2}$ and $\ket{3z^2-r^2}$ in the (${Q_2}$, ${Q_3}$) space as given by \cite{zhou2006unusual,zhou2008orbital}\\
\begin{eqnarray*}
 \psi(\theta) = cos(\theta/2)\ket{3z^2-r^2}+sin(\theta/2)\ket{x^2-y^2}
 \end{eqnarray*}
where the angle $\theta$ ($\theta$ = 90$^0$ +$ |\phi|$) represents respective orbital components, which opens anticlockwise from the ${Q_3}$ axis. The $\theta$ = 0, 2$\pi$/3 and 4$\pi$/3 correspond to orbitals $\ket{3z^2-r^2}$, $\ket{3y^2-r^2}$ and $\ket{3x^2-r^2}$, respectively and $\theta$ = $\pi$/3, $\pi$ and 5$\pi$/3 represent $\ket{y^2-z^2}$, $\ket{x^2-y^2}$ and $\ket{z^2-x^2}$, respectively. An octahedral site distortion, which has a $\theta$ deviating from these special angles reflects either the presence of orthorhombic distortion or a combination of orbital ordering and orthorhombic distortion. Figure 3 (b) depicts the schematic representation of orbital ordering in Mn$^{3+}$ ions in GdMnO$_3$. Since for all compositions in GdMn$_{1-x}$Cr$_x$O$_3$, $\theta$ falls between the special angle and close to 2$\pi$/3 for one of the co-planer Mn-sites (site 1) as defined in Fig. 2 (e) (it is close to 4$\pi$/3 for site 2), a new angle $\gamma$ = $\pi$/6 - $\phi$ can be defined to simplify the wave functions for occupied ($e_g^1$) and unoccupied ($e_g^2$) orbitals for site 1, such that \cite{zhou2006unusual,zhou2008orbital}\\
\begin{eqnarray*}
 \psi_{occ}(\gamma) = cos(\gamma/2)\ket{3y^2-r^2} + sin(\gamma/2)\ket{z^2-x^2}\\
\psi_{unocc}(\gamma) = -sin(\gamma/2)\ket{3y^2-r^2} + cos(\gamma/2)\ket{z^2-x^2}
\end{eqnarray*}
\begin{figure}[!ht]
 \centering
 \includegraphics[height=6cm,width=8cm]{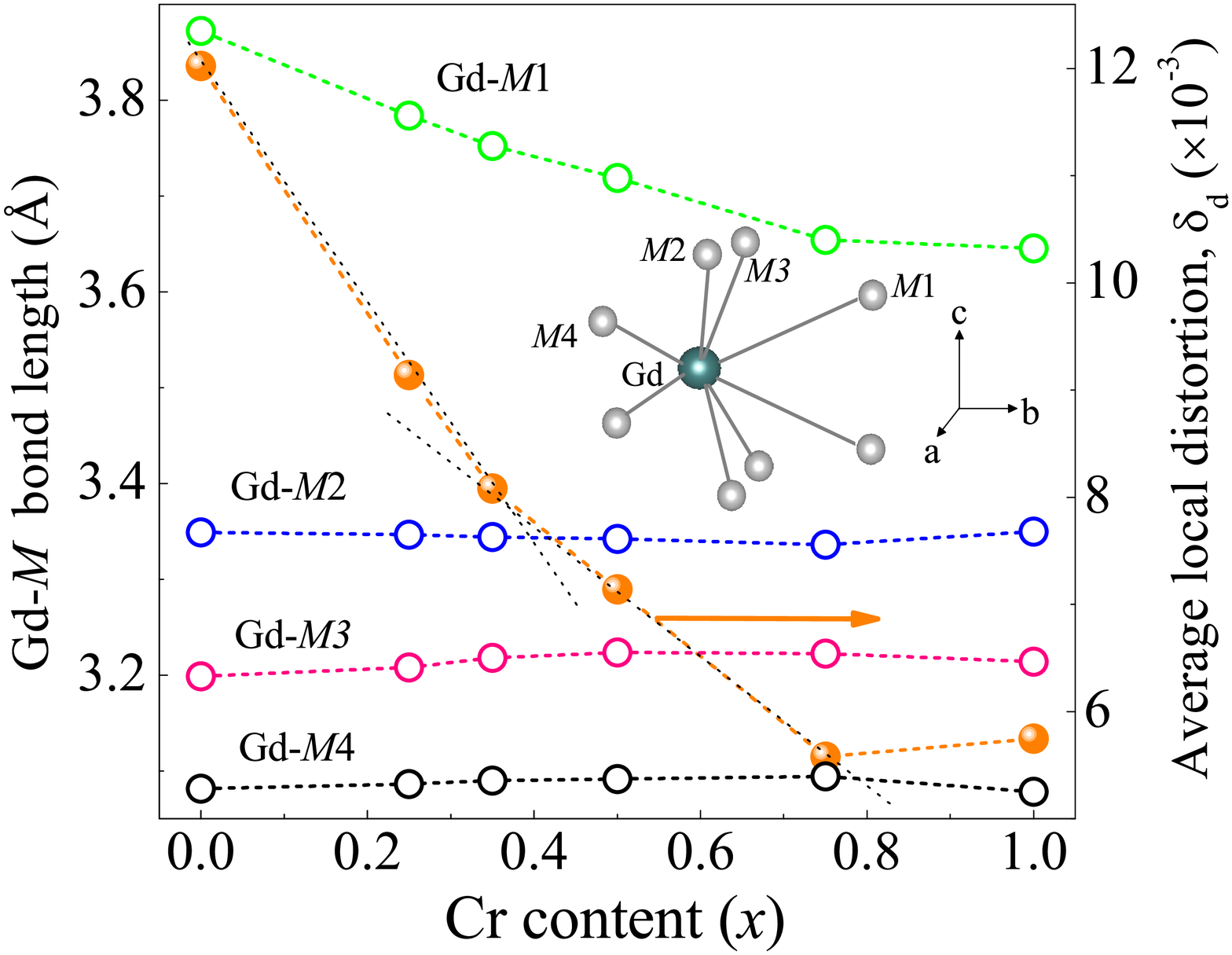}
\caption{(Color online) The evolution of the lattice anisotropy specific to the local Gd environment through variations in the nearest-neighbor Gd-$M$ bond lengths. Right panel represents corresponding average distortion ($\delta_d$) in the local Gd-$M$ environment. Dotted lines guiding $\delta_d$ signify the  slope crossover around $x$ $\sim$ 0.35. Inset represents nearest-neighbor Gd-$M$ bond lengths with blue atom,Gd and gray atoms, $M$. All the parameters are derived from the Rietveld refinements of the respective XRD patterns.}
 \label{fig3}
 \end{figure}
The total site distortion, as measured by $\rho_0$, remains above 0.3 up to $x$ $\sim$ 0.35 compositions, reflecting a dominant contribution from a static JT orbital mixing along with the octahedral distortion \cite{zhou2006unusual,zhou2008orbital}. For $x$ = 0.5 and higher compositions, $\rho_0$  is about one order of magnitude smaller than that of JT-active GdMnO$_3$ and other manganites ($R$MnO$_3$) and resembles the one found in JT-inactive rare-earth ferrites ($R$FeO$_3$) and vanadites ($R$VO$_3$) indicating the disappearance of  orbital ordering \cite{zhou2006unusual,zhou2008orbital,zhou2008intrinsic}. \\

 To examine lattice anisotropy specific to the local Gd environment, nearest-neighbor Gd-$M$ bond lengths are plotted, as shown in Fig. 4. There are eight NN-coordinated $M$ which are doubly paired as $M$1-$M$4, as viewed schematically in the inset of Fig. 3. Each pair of Gd-$M$ lengths are equivalent for all compositions except GdCrO$_3$, in which they are unequal (slightly) due to the ${Pna2_1}$ symmetry. The longest Gd-$M$1(shortest Gd-$M$4) lying in the ${bc}$-plane, shows visible reduction (slight increase) towards $x$ = 1, which is due to the suppression of JT distortion predominantly along the $b$-axis. The corresponding average local distortion, $\delta$$_d$ [= (1/8) $\sum$$_{n=1-8}$ [(${d_n}$-$\braket{d}$)/$\braket{d}$]$^2$, where ${d_n}$ ($\braket{d}$) is the individual (average) Gd-$M$ bond length] also shows a decrease of local anisotropy with increase of Cr-content followed by a slight increase in GdCrO$_3$ owing to having ${Pna2_1}$ symmetry. A slope change occurs (guided by dotted line) around the critical concentration, $x$ $\sim$ 0.35 of JT-crossover, consistent with earlier discussions. This suggests that the evolution of lattice and orbital degrees freedom in the solid solutions is also correlated with the ${bc}$-plane anisotropy in the local Gd-environment \cite{chiang2011effect}.\\
 \begin{figure}[!ht]
 \centering
 \includegraphics[height=11cm,width=8cm]{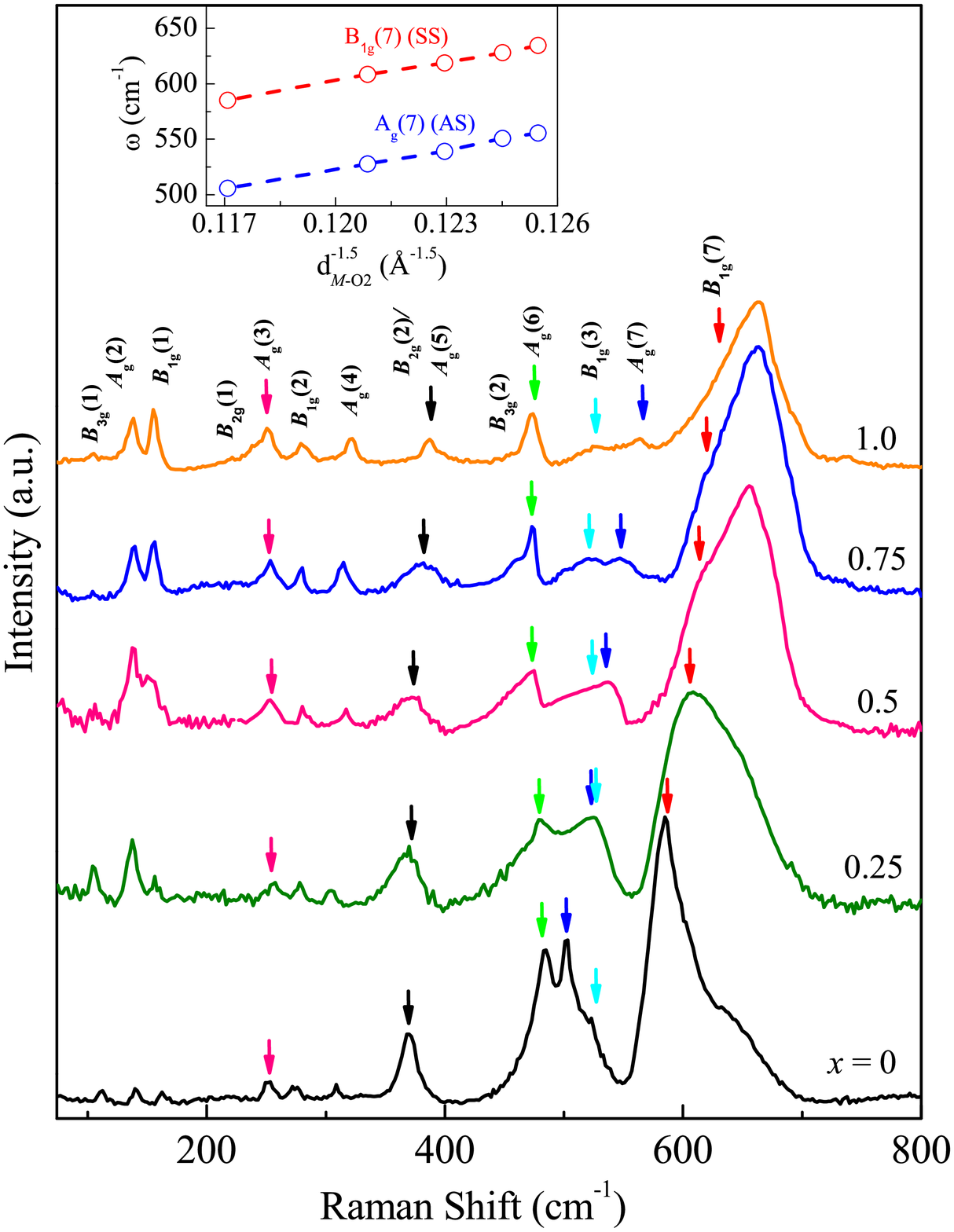}
\caption{(Color online) (a) Evolution of room temperature Raman spectra with compositions ($x$) in GdMn$_{1-x}$Cr$_x$O$_3$ ($\emph{x}$ = 0, 0.25, 0.5, 0.75 and 1.0). The inset shows the linear dependence of the JT symmetric stretching [$B_{2g}$(7)] and antisymmetric stretching [$A_{g}$(7)] modes frequency with the $d$$^{-1.5}_{M-O2}$ , where $d$$_{M-O2}$ is the average of short and long $M$-O2 bond lengths.}
 \label{fig4}
 \end{figure}
In order to understand the lattice/atomic vibrations present in the above system and their role in the structural deformations, room temperature Raman spectroscopy measurements were performed in the solid solutions, as depicted in Fig. 4. For isostructural orthorhombic structure, the group theory predicts 24 Raman active modes (7$A_g$ + 7$B_{1g}$ + 5$B_{2g}$ + 5$B_{3g}$) at the $\Gamma$ point of the Brillouin zone \cite{mahana2017local,iliev2006distortion,todorov2011comparative}, where as only 14 Raman active modes were observed. This may be because of the fact that the other predicted modes are either too low in intensity or beyond our experimental range to be observed. The details about the observed modes are described elsewhere\cite{mahana2017local, staruch2014evidence,iliev2006distortion,todorov2011comparative}. The modes around 670 cm$^{-1}$ may be the disorder-induced phonon density of states of oxygen vibrations \cite{iliev2007multiple,iliev2003role}. Apart from this, Kovaleva ${et}$ ${al.}$ argued that there is an additional component to the multi-order scattering, may arise from coupling between the low-energy electronic motion and the vibrational modes \cite{kovaleva2013anomalous}.

\begin{figure*}[!ht]
 \centering
 \includegraphics[height=12cm,width=17cm]{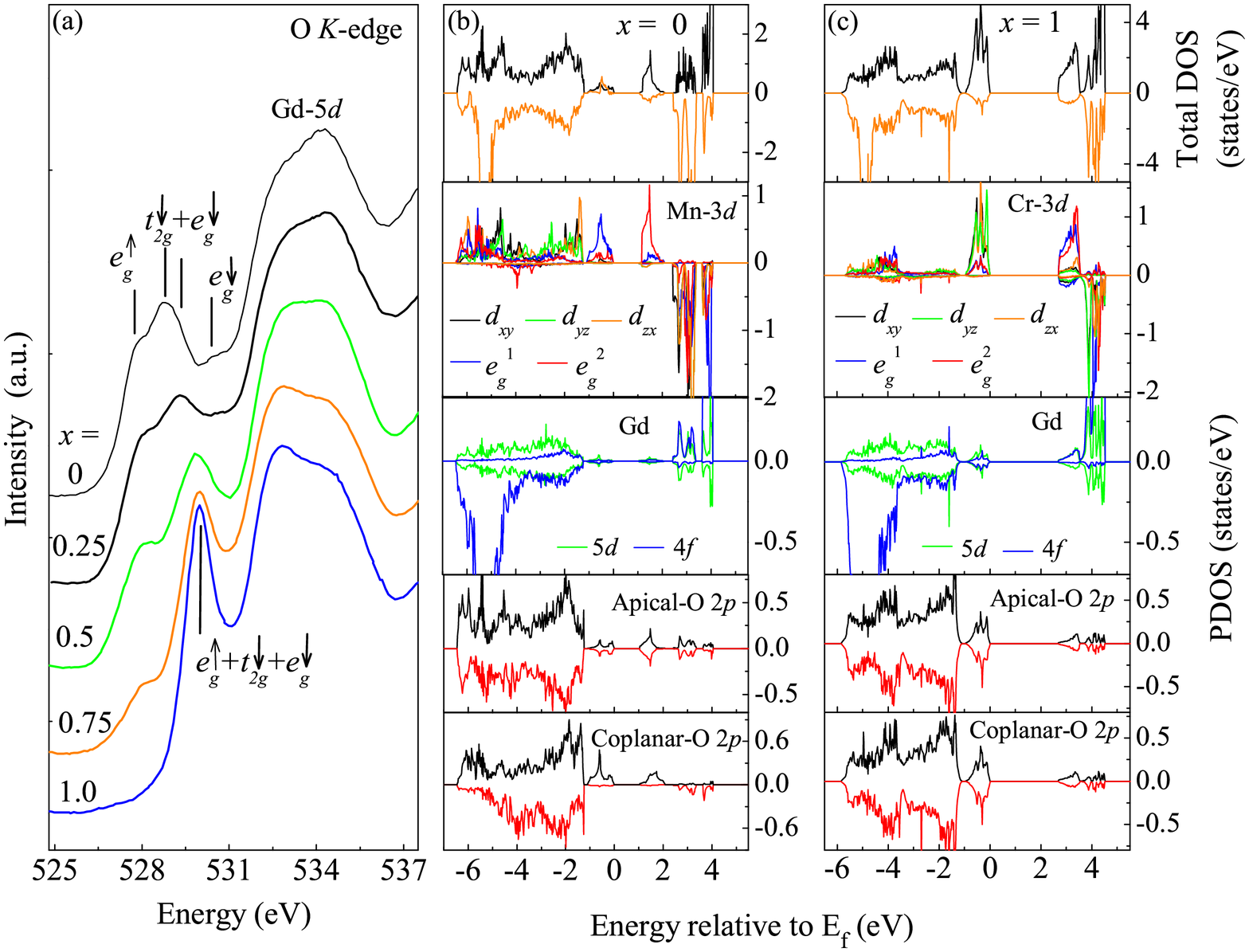}
\caption{(Color online) (a) Evolution of O $K$-edge XAS spectra with compositions ($x$) in GdMn$_{1-x}$Cr$_x$O$_3$ ($\emph{x}$ = 0, 0.25, 0.5, 0.75 and 1.0). (b) and (c) The total and site-decomposed DOS of GdMnO$_3$ and GdCrO$_3$, respectively.}
 \label{fig5}
 \end{figure*}
The most common distortion in orthorhombic ${Pbnm}$ stucture is the tilting of $B$O$_6$ octahedra, which can be described either by orthogonal tilt angles (leading to the ${a^-a^-c^+}$ Glazer’s notation) or by octahedral tilts $\theta$, $\phi$ and $\Phi$ around the pseudocubic [110]$_{pc}$ and [001]$_{pc}$  and [111]$_{pc}$ axis \cite{glazer1975simple,zhao1993critical,weber2012phonon,daniels2013structures}. The Raman modes, $A_g$(3) and $A_g$(5) are correlated with the tilt angles $\phi$ and $\theta$, respectively \cite{weber2012phonon,iliev2006distortion}. The position of $A_g$(3) remains unchanged throughout the series i.e. $\phi$ remains constant. In contrast, $A_g$(5) shows hardly any shift up to $x$ $\sim$ 0.5 compositions and thereafter, it shifts around 15 cm$^{-1}$ towards high frequency for $x$ = 0.75 and 1.0 compositions. This, in turn, suggests that $\theta$ remains more or less constant up to $x$ $\sim$ 0.5 compositions and slightly increase (negligibly small) for $x$ = 0.75 and 1.0 compositions. Furthermore, $\Phi$ is correlated with $\theta$ and $\phi$ via the relation, cos$\Phi$ = cos$\theta$ cos$\phi$ \cite{zhao1993critical,weber2012phonon}, implying that $\Phi$ also increases slightly for $x$ = 0.75 and 1.0 compositions. The in-plne antisymmetric stretching [$A_g$(7)] and symmetric stretching [$B_{1g}$ (7)] modes are the JT modes, which are associated with the $M$-O2 bond lengths in ${ab}$-plane. The variation of frequency of these modes follows the relation, $\omega$ $\propto$ $d^{-1.5}_{M-O2}$ \cite{iliev2006distortion,martin2002raman}, as shown in the inset of Fig. 4. $A_g$(7) and $B_{1g}$ (7) modes show a clear shift up to $x$ $\sim$ 0.5 composition. Previously parameters extracted from the XRD patterns, depicted in Fig. 2,  shows a rapid decrease of $M$-O2 bond length with Cr doping up to $x$ $\sim$ 0.5, suggesting that the clear shift in modes are arising from the rapid decrease of $M$-O2 bond length. Beyond $x$ $\sim$ 0.5 composition, both the variation in $M$-O2 bond length as well as the shift in modes become lesser suggesting there strong interdependence. With Cr-doping there is a rapid decrease of $M$-O2 bond length up to $x$ $\sim$ 0.5 (Fig. 2) leading to clear shift of these two modes after that shift is less. Furthermore, the spectral weight also decreases dramatically with increase of Cr-content ($x$) due to the reduction of JT distortion and becomes weak for .75 and 1.0 compositions, which are the JT-inactive compounds.  \\
 X-ray absorption spectroscopy (XAS) measurements were performed at the O $K$-edge of the solid solutions for $x$ = 0, 0.25, 0.5, 0.75 and 1 to obtain   information about the unoccupied ${M}$ 3${d}$, Gd 5$d$ and deep Gd/${M}$ states via the hybridization with the O 2$p$ states \cite{chiang2011effect,chen2009bonding}, see Fig. 5 (a). Density of state (DOS) calculations were performed on the end compositions GdMnO$_3$ ($x$ = 0) and GdCrO$_3$ ($x$ = 1) and their total and site decomposed DOS are shown in Fig. 5 (b) and (c), respectively. Due to JT orbital ordering, the $e_g$$\uparrow$ band splits into two sub-bands: occupied $e_g^1$$\uparrow$ mainly dominated by $\ket{3y^2-r^2}$ with a small contribution from $\ket{z^2-x^2}$ and unoccupied $e_g^2$$\uparrow$, mainly contributed by $\ket{z^2-x^2}$ mixing with $\ket{3y^2-r^2}$ for one of the co-planar Mn sites (for other co-planar Mn sites, these two $e_g$ orbitals are a mixture of  $\ket{3x^2-r^2}$ and $\ket{y^2-z^2}$), as discussed earlier. Based on the DOS calculations, it can be deciphered that the features between 526.5-531 eV in the O $K$-edge spectra of GdMnO$_3$ are contributed by unoccupied $e_g$$\uparrow$ ($e_g^2$$\uparrow$), $t_{2g}$$\downarrow$+$e_g$$\downarrow$ and $e_g$$\downarrow$\cite{chiang2011effect,chen2009bonding,chen2010strong}, as labeled in Fig. 5 (a). The first peak around 527.8 eV arises from $e_g^2$$\uparrow$ states. From the partial density of state (Fig. 6(b)), a band gap of 1.2 eV is obtained between these JT-split Mn $e_g$$\uparrow$ bands, which is in agreement with the calculated indirect band gap from absorption study of polycrystalline GdMnO$_3$ \cite{bukhari2016infrared}. The second peak of XAS spectrum around 529 eV is associated with $t_{2g}$$\downarrow$ and $e_g$$\downarrow$ states, with a small contribution from $e_g$$\downarrow$ around 530.5 eV. Further there is a splitting of 1.4 eV between $e_g^2$$\uparrow$, $t_{2g}$$\downarrow$+$e_g$$\downarrow$ states, which agrees satisfactorily with the calculations. \\

In contrast to GdMnO$_3$, the O $K$-edge for the JT-inactive GdCrO$_3$ shows a single peak between 529-531 eV contributed by all unoccupied states of Cr$^{3+}$ i.e.  $e_g$$\uparrow$, $t_{2g}$$\downarrow$ and $e_g$$\downarrow$ states and is in good agreement with the calculated PDOS (Fig. 5 (c)). The energy gap of  2.7 eV obtained from the calculation agrees well with the experimentally obtained band gap values reported in chromite family \cite{kotnana2015band,gupta2016study}. Above studies suggest that the first hump (527.8 eV) in XAS spectra for intermediate compositions generally arise from contribution of $e_g^2$$\uparrow$ state of Mn-atoms and second broad hump contributes from $e_g$$\uparrow$ state of Cr and $t_{2g}$$\downarrow$ and $e_g$$\downarrow$ state of both Mn and Cr ions. A gradual decrease of the first peak with increasing of Cr composition is clearly observed. \\
\begin{figure*}[!ht]
 \centering
 \includegraphics[height=9cm,width=12cm]{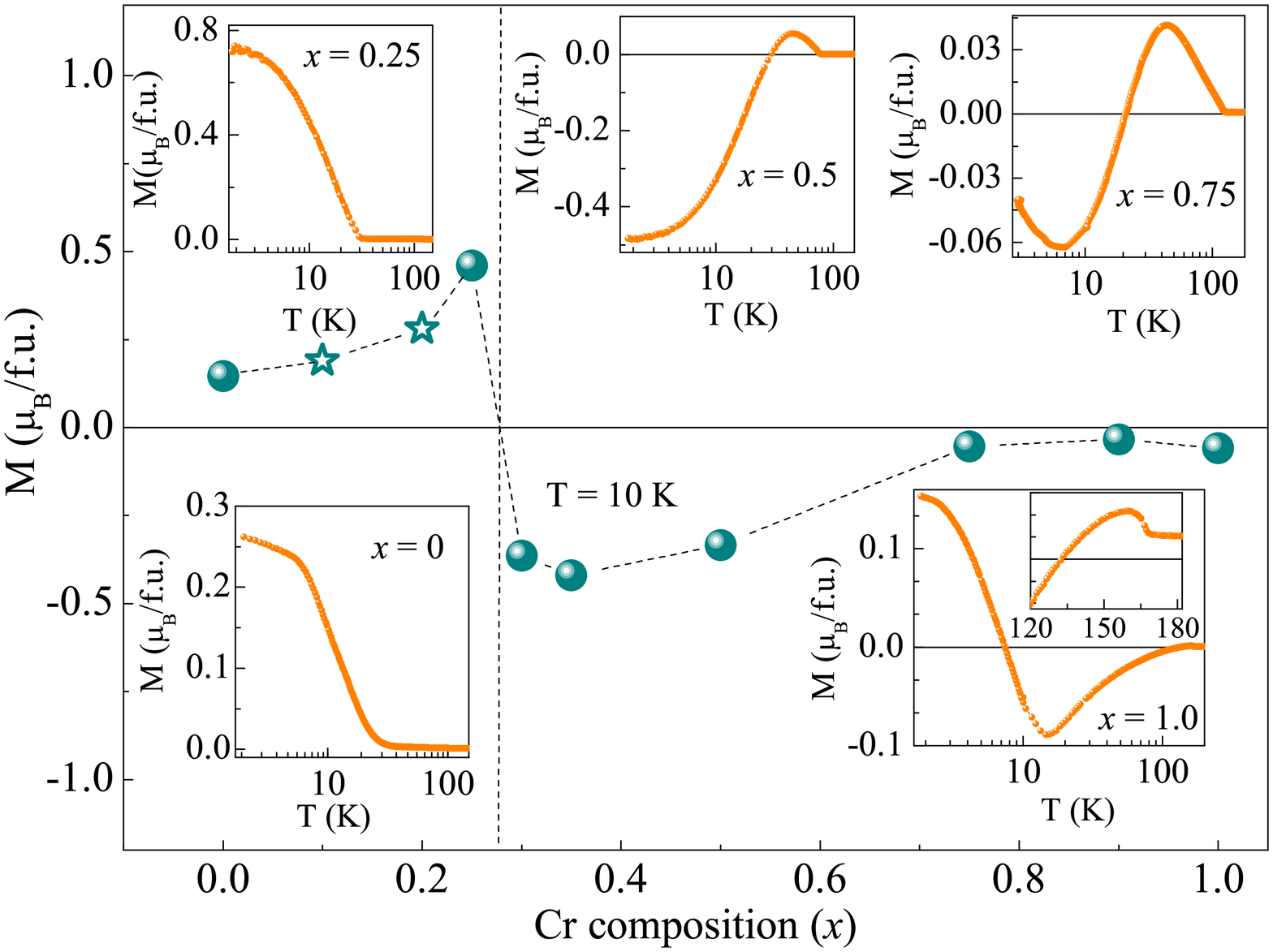}
\caption{(Color online) Variation in magnetic moment at 10 K with compositions ($x$) in GdMn$_{1-x}$Cr$_x$O$_3$.  Stars ($\ast$) represent the results extracted from Ref. 21. Insets represent temperature dependent magnetization measured in field-cooled-cooling protocol of the solid solutions for $x$ = 0, 0.25, 0.5, 0.75 and 1.}
 \label{fig6}
 \end{figure*}
The broad feature between 531-537 eV  (Fig. 5 (a)) corresponds to Gd 5$d$ states, indicating hybridization of Gd 5$d$ and $M$ 3$d$ ($e_g$$\uparrow$) states through the mediation of oxygen, suggesting Gd(4$f$)-$M$(3$d$) electronic interactions in these compounds \cite{richter1998band,stroppa2010multiferroic}. Owing to the highly localized character of the 4$f$ states, direct 3$d$($M$)-4$f$(Gd) coupling is unlikely. Hence the interactions occur via 5$d$ states as they are partially polarized by the 4$f$ electrons via intra-atomic 4$f$-5$d$ exchange interactions and finally couples with the $M$-3$d$ states mediated by the O-2$p$ states \cite{richter1998band,stroppa2010multiferroic}. The 3$d$-4$f$ interactions via hybridized 5$d$ and O-2$p$ states is also evidenced from the calculated DOS of GdMnO$_3$ and GdCrO$_3$ (Fig. 5 (b) and (c)).
\begin{table}[h!]
\centering
\caption{Calculated relative energies ($E$, in meV/unit cell) of various magnetic structures of GdMn$_{0.5}$Cr$_{0.5}$O$_3$. The unit cell contains two Mn and two Cr spins. The energies of the FM phase with layer-by-layer arrangements is used as the reference energy. Subscripts m and 'c' represent Mn and Cr ions, respectively.}
\begin{tabular}{|m{1.5cm}| m{1.2cm}|m{1.2cm}|m{1.2cm}|m{1.2cm}|m{1.2cm}|}
      \hline
    Magnetic  &c$\uparrow$ c$\uparrow$& c$\downarrow$ c$\uparrow$& c$\downarrow$ c$\downarrow$&c$\uparrow$ c$\downarrow$ &c$\uparrow$ c$\downarrow$\\ [1ex]
   structure&m$\uparrow$ m$\uparrow$&m$\uparrow$ m$\uparrow$&m$\uparrow$ m$\uparrow$&m$\downarrow$ m$\uparrow$&m$\uparrow$ m$\downarrow$\\ [1ex]
  \hline
  $E$ & 0 &-16.15&12.25&131.02&125.33 \\[1ex]
  \hline
Magnetic  &m$\uparrow$ c$\uparrow$& m$\uparrow$ c$\downarrow$& m$\downarrow$ c$\uparrow$&m$\downarrow$ c$\downarrow$ &\\ [1ex]
   structure&m$\uparrow$ c$\uparrow$&m$\uparrow$ c$\uparrow$&m$\uparrow$ c$\uparrow$&m$\uparrow$ c$\uparrow$&\\ [1ex]
  \hline
 $E$ & 354.87 &332.98&420.76&421.31& \\[1ex]
  \hline
Magnetic  &c$\uparrow$ m$\uparrow$& c$\downarrow$ m$\downarrow$& c$\uparrow$ m$\downarrow$&c$\downarrow$ m$\uparrow$& \\ [1ex]
   structure&m$\uparrow$ c$\uparrow$&m$\uparrow$ c$\uparrow$&m$\uparrow$ c$\uparrow$&m$\uparrow$ c$\downarrow$&\\ [1ex]
  \hline
 $E$ & 321.31 &306.16&343.92&372.3& \\[1ex]
  \hline
  \end{tabular}
   \label{tab:example}
\end{table}
Temperature dependent magnetization measurements were performed on the solid solutions, GdMn$_{1-x}$Cr$_x$O$_3$, for 0 $\le$ $x$ $\le$ 1. Figure 6 depicts variation in magnetic moment at 10 K with doping concentration and insets represent temperature dependent magnetization measured in field-cooled-cooling (FCC) mode for various $x$ values as indicated. Due to the co-operative JT orbital ordering in GdMnO$_3$, NNN-AFM coupling plays a significant role in addition to NN-FM coupling in the ${ab}$-plane.  There exists an AFM coupling between these layers along the $c$-axis. The competition between these leads to sinusoidal ordering below 40 K followed by canted $A$-type (${A_yF_z}$ in Bertaut's notation) ordering below 20 K in Mn-sublattice as described in detail in our earlier report \cite{mahana2017giant}. Remarkably, magnetization at low temperature (10 K) increases gradually upon Cr-doping up to $x$ $\sim$ 0.25 in spite of the fact that Cr$^{3+}$ moment is smaller than the Mn$^{3+}$ moment, indicating the strengthening of FM interactions in the system. This is probably due to the increase in the strength of  NN-FM coupling as compared to NNN-AFM coupling in Mn-sublattice caused by progressive decrease of JT distortion. Furthermore, possibility of having magnetic interactions, probably FM in nature, between the Mn$^{3+}$ and Cr$^{3+}$ ions, can not be ignored, since such Mn$^{3+}$-Cr$^{3+}$ FM coupling are reported previously in other Mn-Cr systems like TbMn$_{1-x}$Cr$_x$O$_3$ \cite{staruch2014evidence}, LaMn$_{1-x}$Cr$_x$O$_3$ \cite{bents1957neutron} and YMn$_{1-x}$Cr$_x$O$_3$ \cite{li2012modulated} systems. The interaction between two Cr$^{3+}$ moments may be ignored in this doping regime because of low Cr concentration. In  GdCrO$_3$, Cr$^{3+}$ ions have $t^3$$e^0$ cubic-field $d$-electron configurations, which leads to an isotropic ${t^3}$-O-${t^3}$ AFM interactions resulting in canted $G$-type ordering (${G_xF_z}$ in Bertaut's notation) in Cr-sublattice below 169 K. The detailed magnetic interactions in GdCrO$_3$ is reported elsewhere \cite{mahana2017complex}. In Cr-rich compositions Cr$^{3+}$-Cr$^{3+}$ interactions dominate, thus have similar behavior to that of GdCrO$_3$ having a $G$-type magnetic structure. The canted spin structures in these systems is a direct consequence of antisymmetric Dzyaloshinskii-Moriya (DM) interaction [$D$. ($\vec{S}$$_i$ $\times$ $\vec{S}$$_j$] \cite{moriya1960anisotropic}. Notably, FCC magnetization curve for $x$ = 0 and 0.25 compositions show positive magnetization in the entire temperature region. However, one sees a magnetization reversal effect for all other compositions suggesting the strengthening of AFM coupling between Gd and $M$ sublattices with increasing Cr-content. This magnetization reversal as a function of temperature above a critical Cr concentration suggests that there is a strong correlation between the structural distortion and magnetic coupling.\\
\begin{figure}[!ht]
 \centering
 \includegraphics[height=8cm,width=10cm]{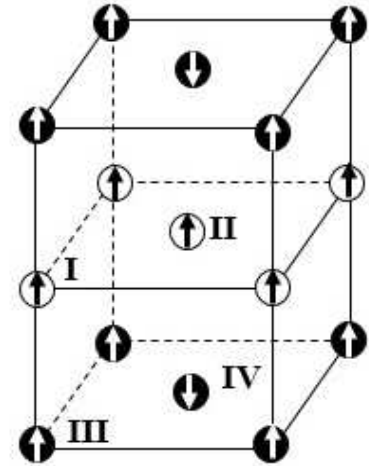}
\caption{(Color online) The calculated most stable magnetic structures in one unit cell for GdMn$_{0.5}$Cr$_{0.5}$O$_3$. Only the transition-metal
ions Mn and Cr are shown (Filled circles: Mn; Empty circles: Cr). The ions labeled as I, II, III and IV are the non-equivalent atoms in the unit cell.}
 \label{fig7}
 \end{figure}
\begin{figure}[!ht]
 \centering
 \includegraphics[height=10cm,width=8cm]{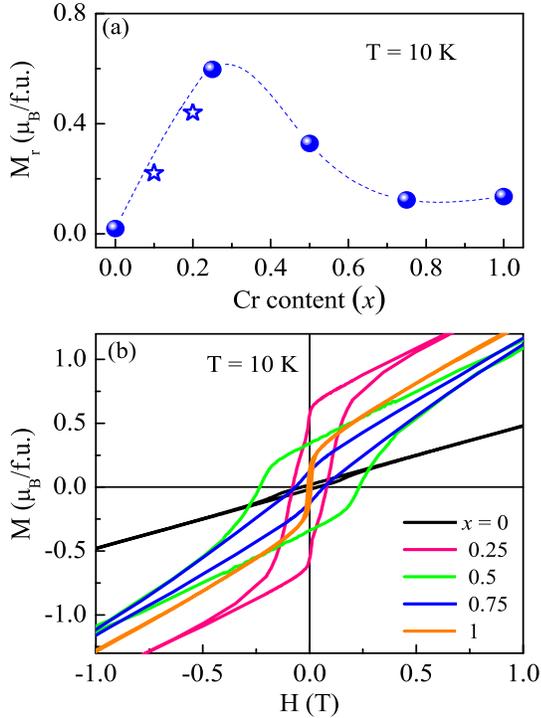}
\caption{(Color online) (a) Variation of remnant magnetization ($M_r$) at 10 K with compositions ($x$). Stars ($\ast$) represent the results extracted from Ref. 21 (b) Enlarged view of $M$-$H$ loops measured at 10 K of the solid solutions for $x$ = 0, 0.25, 0.5, 0.75 and 1.}
 \label{fig8}
 \end{figure}
DFT calculations were performed to determine the possible low temperature magnetic ground state in GdMn$_{0.5}$Cr$_{0.5}$O$_3$. Previously reports on similar Mn-Cr systems with composition $R$Mn$_{0.5}$Cr$_{0.5}$O$_3$ suggest that the magnetic interactions in these systems strongly depend on the $R$-sites. For e.g., TbMn$_{0.5}$Cr$_{0.5}$O$_3$ exhibits $G$-type magnetic structure with the alternate arrangements of Mn and Cr atoms as confirmed from neutron diffraction study and DFT calculations \cite{staruch2014magnetic}. DyMn$_{0.5}$Cr$_{0.5}$O$_3$ has random distributions of Mn and Cr ions, which leads to two distinct magnetic orderings associated with Cr$^{3+}$-Cr$^{3+}$ and Cr$^{3+}$-Mn$^{3+}$ interactions as clearly seen in temperature dependent magnetization data \cite{yuan2015dielectric}. From DFT calculations, LaMn$_{0.5}$Cr$_{0.5}$O$_3$ was found to have stable structure with the layer-by-layer doping type (Mn and Cr are alternatively arranged along the $c$-axis). The DFT calculations also showed FM interactions between Mn ions and AFM interactions between Cr ions in the ${ab}$-plane and satisfactorily explained its magnetization \cite{yang2000density}. All the compounds discussed above possess ${Pbnm}$ symmetry. On the contrary, YMn$_{0.5}$Cr$_{0.5}$O$_3$ has monoclinic structure with layer-by-layer arrangements of Mn and Cr along the $c$-axis and exhibits ferrimagnetic behavior \cite{yang2013ferrimagnetism,hao2014layered}. In the present system GdMn$_{0.5}$Cr$_{0.5}$O$_3$, Reitveld refinement of XRD pattern using orthorhombic ${Pbnm}$ structure showed better fitting  than that of the monoclinic structure. Also temperature dependent magnetization measurements did not show double transitions like in DyMn$_{0.5}$Cr$_{0.5}$O$_3$, implying that Mn and Cr are well ordered. \\

To understand the magnetic coupling in GdMn$_{0.5}$Cr$_{0.5}$O$_3$, the total energy was calculated within the framework of GGA including Hubbard $U$, for various possible arrangements of Mn and Cr ions and various possible spin configurations, as listed in Table I. The results of GGA+$U$ are consistent with the results including spin-orbital coupling (SOC) i.e. GGA+$U$+SOC as reported in TbMn$_{0.5}$Cr$_{0.5}$O$_3$ \cite{staruch2014magnetic}. It is found that structure with the layer-by-layer doping type with a FM interaction between Mn$^{3+}$ spins and AFM interaction between Cr$^{3+}$ spins is the most stable configuration similar to that of LaMn$_{0.5}$Cr$_{0.5}$O$_3$ \cite{yang2000density}.\\

In constructing an effective Spin Hamiltonian (SH) to understand the spin dependent energetics of   GdMn$_{0.5}$Cr$_{0.5}$O$_3$ only Mn$^{3+}$ and Cr$^{3+}$ spins were considered , since Gd$^{3+}$ spins were not included in the total energy calculations. The unit cell used in the calculation of the energies for different Mn and Cr arrangements and different spin orientations consists of 20 atoms, Gd$_4$Mn$_2$Cr$_2$O$_{12}$, consisting of two Mn and two Cr ions.  The unit cell then consists of 4 magnetic atoms I-IV; I, II are in one basal plane representing the Mn atoms and III, IV are in the other basal plane representing the Cr atoms as shown schematically in Fig. 7. The structure in the figure is denoted as (Mn$\uparrow$,Mn$\uparrow$)(Cr$\uparrow$,Cr$\downarrow$), which is the lowest energy atomic structure obtained from the calculations. \\

The SH for the system is given by 
\begin{eqnarray}
 H_{spin}=-J_1\sum_{<ij>l}{\vec{S}_{il}^{Mn}.{\vec{S}_{jl}^{Mn} }}-J_2\sum_{<ij>l}{\vec{S}_{il}^{Cr}.{\vec{S}_{jl}^{Cr}}}\nonumber \\
-J_3\sum_{<ll^\prime>l}{\vec{S}_{il}^{Mn}.{\vec{S}_{il^\prime}^{Cr} }}+\Delta
\end{eqnarray}
where $i$, $j$ indicate lattice sites in the basal (${ab}$) plane and $l$ indicates different layers along the $c$ axis. The number of Mn-Mn bonds Cr-Cr bonds and Mn-Cr bonds per unit cell are four each, as presented in Fig. 8. There are four parameters in the SH which were estimated using calculated DFT energies for different spin configurations with layer-by-layer arrangements of Mn and Cr along the $c$-axis (see Table I first row). Here S$^{Mn}$=2; S$^{Cr}$=3/2. In the mean-field approximation, energy/unit-cell (in mev/unit cell) for four different spin configurations are given below. Energy for a fifth configuration can be predicted within SH model and compared with DFT energy.\\

\begin{eqnarray}
 E_I= -16J_1-9J_2-12J_3+\Delta=0
\end{eqnarray}
\begin{eqnarray}
E_{II}=-16J_1+9J_2+\Delta=-16.15
\end{eqnarray}
\begin{eqnarray}
E_{III}=-16J_1-9J_2+12J_3+\Delta=+12.25
\end{eqnarray}
\begin{eqnarray}
E_{IV}=+16J_1+9J_2+12J_3+\Delta=+131.02
\end{eqnarray}
\begin{eqnarray}
E_V=+16J_1+ 9J_2-12J_3+\Delta
\end{eqnarray}
The Eqs. 2-5 can be solved to give: $J_1$=$J^{Mn-Mn}$ = 4.39 meV (Ferro), $J_2$ = $J^{Cr-Cr}$ = -1.26 meV (Antiferro), $J_3$ = $J^{Mn-Cr}$ = 0.55 meV (Ferro) and $\Delta$ = 65.48 meV. Using these parameters E$_V$ = 117.78 meV can be predicted, whereas the calculated DFT energy is 125.33 meV.
Looking at the calculated exchange parameters, the coupling between Mn spins is ferromagnetic and strong, similar to the parent manganite, i.e. GdMnO$_3$. The coupling between Cr spins is antiferromagnetic, again similar to the parent compound GdCrO$_3$. The cross coupling between Mn and Cr spin is ferromagnetic and weak. Earlier report on transport studies in LaMn$_{1-x}$Cr$_x$O$_3$; 0$\le$ $x$ $\le$ 0.15. argues that the experimental findings could be understood through a simple model of the electronic structure of the alloy, which included FM double exchange interaction between Mn and Cr spins and a CPA type approximation to handle disorder \cite{PhysRevB.72.132413}. It is possible that the small magnitude of $J^{Mn-Cr}$ results from a near cancellation between two competing contributions, one the usual AFM super-exchange and the other FM double exchange.\\

 Figure 8 (a) and (b) shows the variation of the remnant magnetization ($M_r$) at 10 K for GdMn$_{1-x}$Cr$_x$O$_3$ for various $x$ values and their corresponding $M$-$H$ loops are shown in , respectively. $M_r$ shows a non-monotonic variation with composition: initially it increases with increasing Cr concentration ($x$) and reaches a maximum value for $x$ $\sim$ 0.3. Beyond $x$ $\sim$ 0.3,  $M_r$ starts decreasing for increasing in $x$ followed by no distinct variation beyond $x$ $\sim$ 0.7. As discussed previously, magnetic behavior in the solid solution is a combination of variety of magnetic interactions such as symmetric exchange interactions (FM and AFM type) and antisymmetric DM interaction coupled to octahedral tilting. The DM interaction is directly proportional to the perpendicular displacement of oxygen in the $M$-O-$M$ chain, which, in turn, depends on the tilt angles \cite{bellaiche2012simple,singh2014magnetic}. It is evident from Raman spectroscopy data that there is hardly any change of tilt angles throughout the series, suggesting that the contributions from canted ferromagnetism (DM interaction) remains almost constant throughout the series. Thus, the increase of  $M_r$ up to $x$ $\sim$ 0.3 suggests an increase of NN-FM coupling as compared to NNN-AFM coupling caused by the progressive decrease of JT-orbital ordering and incorporation of Mn$^{3+}$-Cr$^{3+}$ FM interactions as discussed earlier. Beyond $x$ $\sim$ 0.3 ferrimagnetic type structure arises due to the incorporation of AFM Cr$^{3+}$-Cr$^{3+}$ interactions, thus resulting in decrease of  $M_r$. In Cr-rich compositions ($x$ $\ge$ 0.7), AFM Cr$^{3+}$-Cr$^{3+}$ interactions are more dominating resulting in no further change in $M_r$.

\section[two column]{conclusion}
The structural, electronic and magnetic properties of GdMn$_{1-x}$Cr$_x$O$_3$ were studied. In the structural investigations,
it was found that the JT distortion characteristic to Mn$^{3+}$ results in bond anisotropy and effective orbital ordering for $x$ $\le$ 0.35. A gradual variation of electronic states with doping is also clearly seen in O-$K$ edge x-ray absorption spectra. The temperature dependence of magnetization under the FCC mode shows sign reversal effect for $x$ $\ge$ 0.35, whereas magnetization does not change sign in the JT-active region. The change in magnetic polarity at the critical concentration coinciding with JT-crossover, infers a complex interplay of  magnetic interaction and structural distortion. The nonmonotonic variation of remnant magnetization can be explained in terms of doping induced modification of  symmetric magnetic interactions (FM/AFM type). DFT calculations using GGA+U type exchange correlation potential find that the system with $x$ = 0.5 consists of alternate ferromagnetic Mn layers and antiferromagnetic Cr layers. The strength of the ferromagnetic exchange interaction between nearest neighbor Mn spins is stronger than the NN Cr antiferromagnetic exchange. The exchange interaction between NN Mn and Cr is quite small but ferromagnetic. This is distinctly different  from that observed for both end members GMnO$_3$ and GdCrO$_3$. 
\section{Acknowledgment}
 B. R. would like to thank Prof. P. V. Satyam, IOP, Bhubaneswar, India for computational facilities.\\

\bibliographystyle{apsrev4-1}
$\dagger$ Present address:  Rajdhani College, Baramunda Square, Nayapalli, Bhubaneswar - 751003, Odisha, India.
\bibliography{MnCr}

\date{\today}

\end{document}